\documentclass[fleqn,usenatbib]{mnras}

\usepackage{newtxtext,newtxmath}

\usepackage[T1]{fontenc}

\DeclareRobustCommand{\VAN}[3]{#2}
\let\VANthebibliography\thebibliography
\def\thebibliography{\DeclareRobustCommand{\VAN}[3]{##3}\VANthebibliography}

\usepackage{graphicx}	
\usepackage{amsmath}	

\title[The repeating fast radio burst FRB 180916.J0158+65]{Statistical properties and lensing effect on the repeating fast radio burst FRB 180916.J0158+65}

\author[Y.-B. Wang et al.]{
Yu-Bin Wang$^{1,2}$,
Abdusattar Kurban$^{1,3,4}$,
Xia Zhou$^{1,3,4}$ \thanks{E-mail: zhouxia@xao.ac.cn}, Yun-Wei Yu$^{5}$
and Na Wang$^{1,3,4}$
\\
$^{1}$Xinjiang Astronomical Observatory, Chinese Academy of Sciences, Urumqi 830011, China\\
$^{2}$University of Chinese Academy of Sciences, 19A Yuquan Road, Beijing 100049, China\\
$^{3}$Key Laboratory of Radio Astronomy, Chinese Academy of Sciences, Nanjing 210008, China\\
$^{4}$Xinjiang Key Laboratory of Radio Astrophysics, 150 Science1-Street, Urumqi 830011, China\\
$^{5}$Institute of Astrophysics, Central China Normal University, Wuhan 430079, China
}

\date{Accepted XXX. Received YYY; in original form ZZZ}

\pubyear{2022}

\begin{document}
\label{firstpage}
\pagerange{\pageref{firstpage}--\pageref{lastpage}}
\maketitle

\begin{abstract}
FRB 180916.J0158+65 is a well-known repeating fast radio burst with a period ($16.35~\rm days$) and an active window ($5.0~\rm days$). We give out the statistical results of the dispersion measures and waiting times of bursts of FRB 180916.J0158+65. We find the dispersion measures at the different frequencies show a bimodal distribution. The peaking dispersion measures of the left mode of the bimodal distributions increase with frequency, but the right one is inverse.
The waiting times also present the bimodal distribution, peaking at 0.05622s and 1612.91266s. The peaking time is irrelevant to the properties of bursts, either for the preceding or subsequent burst. By comparing the statistical results with possible theoretical models, we suggest that FRB 180916.J0158+65 suffered from the plasma lensing effects in the propagation path. Moreover, this source may be originated from a highly magnetized neutron star in a high-mass X-ray binary.
\end{abstract}

\begin{keywords}
pulsars: general $-$ stars: individual (FRB 180916.J0158+65) $-$ transients: fast radio bursts $-$ interstellar medium $-$ scattering
\end{keywords}



\section{Introduction}

Fast radio bursts (FRBs) are mysterious bright (emitting $\rm \sim Jy$) radio sources that can release extremely high energy within milliseconds of time scale. Up to now, more than 600 FRBs have been detected, and 24 of them are repeaters \citep{Petroff2016,Luo2020,Amiri2021}.
Their unusually high dispersion measures (DMs) indicate that they are extragalactic or cosmic origin rather than Galactic \citep{Thornton2013}.
Recently, FRB 20200428 from the magnetar SGR 1935+2154 has been detected the similar properties as the repeaters \citep{Andersen2019}, and the bursts of FRB 180301 have the analogous polarization angle swings to pulsars \citep{Luo2020}.
These suggest that the repeating sources could be the possibility of luminous coherent emission processes around pulsars or magnetars \citep{Kumar2017,Andersen2019,Andersen2020,Li2021a}.
However, FRBs are still topical events and keep mysterious in its physical nature.

The repeating FRBs may have two origins, i.e., the isolated neutron stars (NSs) and the NS binary systems \citep{Platts2019,Kurban2022}.
An isolated NS requires the accumulation of enough energy to release the next burst randomly, which presents the propriety of aperiodicity.
Thus, the investigation for waiting time ($\Delta t$), the time interval between two adjacent bursts within an observational campaign, is pretty important for understanding the physical nature of FRBs.
For example, the waiting time with unimodal distribution has been found in the bursts activities from the magnetars in the Milky Way \citep{Cheng2020,Younes2020} or the giant pulses from the isolated pulsar \citep{Abbate2020,Kuiack2020,Geyer2021}.
However, the waiting times of FRB 121102 presented a bimodal distribution, which is uncorrelated with the fluences, peak flux densities, pulse widths, and high energy components of bursts (\citealp{Li2019}; \citealp{Li2021b}).
The waiting times of FRB 20201124A also has a bimodal distribution \citep{Xu2022}.
These indicate that some repeaters, like FRB 121102 and FRB 20201124A, may originate from the activity in the binary systems \citep{Du2021,Geng2021,Wang2022a} rather than the isolated NS.

FRB 180916.J0158+65 exhibited similar observational properties to FRB 121102 and FRB 20201124A. For example, some bursts among the three repeaters present shorter delay times at low frequency than that at high frequency \citep{Chamma2021,Platts2021,Kumar2022}.
In addition, they are located behind the clump of plasma \citep{Tendulkar2017,Marcote2020,Xu2022}; they occupy an optical counterpart \citep{Tendulkar2017,Li2022,Ravi2022}; they have the periodicity and an active window \citep{Amiri2020,Rajwade2020,Mao2022}; and they produce a delay time of about tens of milliseconds between bursts \citep{Amiri2020,Li2021b,Xu2022}.
These indicate FRB 180916.J0158+65 may have similar statistical properties to FRB 121102.
Thus, we give out the statistic of the DM and waiting time of FRB 180916.J0158+65, which is detected at frequencies from 110 to 5.4 GHz.
Coupled with the statistical results for the repeater, the effects in the propagation path (gravitational lensing and plasma lens) will be discussed, which may help us to reveal the physical nature of this source.

This paper is organized as follows. In Section 2, we give out the statistical properties of FRB 180916.J0158+65, especially for waiting time and DMs. The lensing effects of the propagation path are discussed in Section 3. Finally, a summary and discussion are given in Section 4.

\section{Statistical properties of FRB 180916.J0158+65}\label{Sect.2}

FRB 180916.J0158+65 with Galactic longitude $l = 129.7^{\circ}$ and latitude $b = 3.7^{\circ}$ is reported a possible $P_{\rm orb} = 16.35$ day periodicity and a 5.0-day phase window approximately.
The Milky Way DM contribution is $200 \, \rm pc \, cm^{-3}$ (NE2001, \citealt{Cordes2002}) or $325.23 \, \rm pc \, cm^{-3}$  (YMW16, \citealt{Yao2017,Amiri2020}). It is located in a nearby spiral galaxy with redshift $z_d = 0.0337 \pm 0.0002$ and luminosity distance $d_{\rm os} = 149.0 \pm 0.9 \, \rm Mpc$ \citep{Marcote2020}.
The projected distance of the repeating source (roughly $4.7 \, \rm kpc$) is far from the core of the host galaxy.
There is no other comparable large and bright galaxy in its observing field of view, but a young stellar clump with size $\sim \rm 380 \, pc$ around the repeater (the source environment $\sim \rm 30-60 \, pc$) \citep{Marcote2020,Tendulkar2021}.
Over 195 bursts have been detected from this source by different instruments at frequencies ranging from 110 MHz to 5.4 GHz \citep{Amiri2020,Chawla2020,Marcote2020,Marthi2020,Pearlman2020,Pilia2020,Pastor-Marazuela2021,Pleunis2021,Bethapud2022,Mckinven2023}. These instruments are including the 100-m Effelsberg Radio Telescope (Effelsberg; 4-5.4 GHz), the European VLBI Network (EVN; 1636-1764 MHz), the Westerbork Synthesis Radio Telescope (WSRT) with the Apertif Radio Transient System (Apertif; 1220-1520 MHz), the upgraded Giant Metrewave Radio Telescope (uGMRT; 550-750 MHz and 250-500 MHz), the Canadian Hydrogen Intensity Mapping Experiment Fast Radio Burst Project (CHIME/FRB; 400-600 MHz), the Green Bank Telescope (GBT; 300-400 MHz), the Sardinia Radio Telescope (SRT; 296-360 MHz) and the Low-Frequency Array (LOFAR; 110$-$188 MHz).
The pulse widths of these bursts are ranging from a few milliseconds to about one hundred milliseconds \citep{Amiri2020,Marthi2020,Pleunis2021}.

\begin{table*}
  \centering
  \caption{Observational  MJD and DMs of FRB 180916.J0158+65.}\label{tb1}
  \renewcommand{\arraystretch}{0.95}
  \setlength{\tabcolsep}{0.4mm}
  \begin{tabular}{lccllccllccl}
  \hline\hline\noalign{\smallskip}
  MJD           & Telescope &          DM           & Ref    &MJD           & Telescope &          DM           & Ref &MJD           & Telescope &          DM           & Ref     \\
                &           & $\rm pc\, \, cm^{-3}$ &        &              &           & $\rm pc\, \, cm^{-3}$ &     &              &           & $\rm pc\, \, cm^{-3}$ &         \\
  \hline\noalign{\smallskip}
  58477.16185   &CHIME  &   $348.732 \pm 0.01$  & [1]    &58950.58347838&LOFAR  &     $349.09\pm0.04$   & [6]    &59095.01647024&Apertif&     $348.70\pm0.40$   & [6]    \\
  58478.15521   &CHIME  &   $348.791 \pm 0.023$ & [1]    &58951.54162736&LOFAR  &     $349.03\pm0.05$   & [6]    &59095.03083630&Apertif&     $348.71\pm0.27$   & [6]    \\
  58638.71643   &CHIME  &   $348.744 \pm 0.032$ & [1]    &58951.55801455&LOFAR  &     $348.98\pm0.02$   & [6]    &59095.03119917&Apertif&     $348.73\pm0.11$   & [6]    \\
  58639.70561   &CHIME  &   $349.89 \pm 0.15$   & [1]    &58951.58470795&LOFAR  &     $348.99\pm0.03$   & [6]    &59095.04813576&Apertif&     $348.74\pm0.16$   & [6]    \\
  58639.71008   &CHIME  &   $348.813 \pm 0.072$ & [1]    &58951.59135120&LOFAR  &     $348.86\pm0.08$   & [6]    &59095.06242878&Apertif&     $348.64\pm0.26$   & [6]    \\
  58653.09613665&EVN    &    $348.76\pm0.1$     & [2]    &58978.59561357&Apertif&     $348.63\pm0.14$   & [6]    &59095.07525644&Apertif&     $348.59\pm0.34$   & [6]    \\
  58653.11125735&EVN    &    $348.76\pm0.1$     & [2]    &58979.50785914&Apertif&     $348.35\pm0.38$   & [6]    &59095.07913932&Apertif&     $348.10\pm0.23$   & [6]    \\
  58653.14659694&EVN    &    $348.76\pm0.1$     & [2]    &58980.35572077&Apertif&     $348.75\pm0.26$   & [6]    &59095.10211216&Apertif&     $349.06\pm0.23$   & [6]    \\
  58653.27850789&EVN    &    $348.76\pm0.1$     & [2]    &58980.38590828&Apertif&     $349.44\pm0.26$   & [6]    &59095.11289895&Apertif&     $348.50\pm0.28$   & [6]    \\
  58786.32075   &CHIME  &   $348.68 \pm 0.46$   & [1]    &58980.44318898&Apertif&     $350.09\pm0.37$   & [6]    &59095.11989684&Apertif&     $348.78\pm0.14$   & [6]    \\
  58802.25840267&GBT    &     $349.3\pm0.2$     & [3]    &58980.46375074&Apertif&     $347.86\pm0.54$   & [6]    &59095.12368446&Apertif&     $348.93\pm0.23$   & [6]    \\
  58835.17324   &CHIME  &   $349.11 \pm 0.2$    & [1]    &58980.46995949&Apertif&     $347.28\pm0.32$   & [6]    &59095.14075045&Apertif&     $349.14\pm0.51$   & [6]    \\
  58836.16929624&GBT    &     $348.8\pm0.4$     & [3]    &58980.47426015&Apertif&     $349.06\pm0.33$   & [6]    &59095.16236684&Apertif&     $349.29\pm0.26$   & [6]    \\
  58836.16929695&GBT    &     $348.8\pm0.4$     & [3]    &58980.52593337&Apertif&     $348.70\pm0.56$   & [6]    &59095.19030365&Apertif&     $349.68\pm0.52$   & [6]    \\
  58836.16929720&GBT    &     $348.8\pm0.4$     & [3]    &58980.54629988&Apertif&     $348.07\pm0.44$   & [6]    &59096.20840871&Apertif&     $348.43\pm0.54$   & [6]    \\
  58836.17198   &CHIME  &   $350.0527\pm 0.0075$& [1]    &58980.54684270&Apertif&     $349.52\pm0.76$   & [6]    &59111.40643   &CHIME  &   $350.12  \pm 0.34$  & [1]    \\
  58836.17591788&GBT    &     $350.1\pm0.4$     & [3]    &58980.62542392&Apertif&     $348.78\pm0.48$   & [6]    &58477.16185   &CHIME  &   $348.732 \pm 0.01$  & [1]    \\
  58852.13628   &CHIME  &   $349.847 \pm 0.083$ & [1]    &58980.62998094&Apertif&     $348.68\pm0.13$   & [6]    &59143.81778929&Apertif&     $348.89\pm0.19$   & [6]    \\
  58852.13773   &CHIME  &   $349.097 \pm 0.061$ & [1]    &58980.65889322&Apertif&     $348.87\pm0.16$   & [6]    &59190.73164688&Apertif&     $348.76\pm0.19$   & [6]    \\
  58868.07586   &CHIME  &   $348.98  \pm 0.12$  & [1]    &58981.38138907&Apertif&     $350.68\pm0.32$   & [6]    &59191.74125466&Apertif&     $348.87\pm0.19$   & [6]    \\
  58868.08221442&GBT    &     $348.9\pm0.1$     & [3]    &58981.77662   &CHIME  &   $349.507 \pm 0.084$ & [1]    &59225.10428   &CHIME  &   $348.99  \pm 0.13$  & [1]    \\
  58868.08461892&GBT    &     $348.7\pm0.2$     & [3]    &58982.76846813&CHIME  &     $352.6\pm3.2$     & [4]    &59241.06071   &CHIME  &   $348.8435\pm 0.0098$& [1]    \\
  58868.08679636&GBT    &     $348.8\pm0.2$     & [3]    &58982.77157   &CHIME  &   $349.46  \pm 0.57$  & [1]    &59244.04400   &CHIME  &   $348.838\pm 0.071$  & [1]    \\
  58882.04681   &CHIME  &   $349.27  \pm 0.27$  & [1]    &58996.15501128&Apertif&     $348.81\pm0.21$   & [6]    &59244.06229   &CHIME  &   $348.6885\pm 0.002$ & [1]    \\
  58883.02020163&CHIME  &     $370.4\pm1.6$     & [4]    &58996.19203445&Apertif&     $348.79\pm0.21$   & [6]    &59245.05833   &CHIME  &   $349.47  \pm 0.31$  & [1]    \\
  58883.03995   &CHIME  &   $349.37 \pm 0.20$   & [1]    &58996.23898191&Apertif&     $348.68\pm0.14$   & [6]    &59275.96877   &CHIME  &   $349.542 \pm 0.051$ & [1]    \\
  58883.04405   &CHIME  &   $349.725 \pm 0.48$  & [1]    &58996.27129126&Apertif&     $348.68\pm0.23$   & [6]    &59276.96257   &CHIME  &   $348.83  \pm 0.45$  & [1]    \\
  58883.05372   &CHIME  &   $348.840 \pm 0.5$   & [1]    &58996.34499129&Apertif&     $350.23\pm0.84$   & [6]    &59277.96034   &CHIME  &   $348.988 \pm 0.02$  & [1]    \\
  58899.00706   &CHIME  &   $348.73  \pm 0.025$ & [1]    &58996.36224320&Apertif&     $348.78\pm0.44$   & [6]    &59306.89010   &CHIME  &   $349.60  \pm 0.17$  & [1]    \\
  58477.16185   &CHIME  &   $348.732 \pm 0.01$  & [1]    &58996.42810299&Apertif&     $349.47\pm0.29$   & [6]    &59355.74759   &CHIME  &   $348.93  \pm 0.37$  & [1]    \\
  58899.56141184&SRT    &     $349.8\pm0.1$     & [5]    &58996.48015176&Apertif&     $348.63\pm0.25$   & [6]    &59357.74680   &CHIME  &   $348.95  \pm 0.34$  & [1]    \\
  58899.56781756&SRT    &     $349.4\pm0.1$     & [5]    &58996.60480633&Apertif&     $348.97\pm0.28$   & [6]    &59390.65203   &CHIME  &   $348.97  \pm 0.21$  & [1]    \\
  58899.57561573&SRT    &     $350.1\pm0.1$     & [5]    &58996.61583838&Apertif&     $348.81\pm0.43$   & [6]    &59406.60025   &CHIME  &   $349.38  \pm 0.14$  & [1]    \\
  58930.47097294&Apertif&     $348.70\pm0.20$   & [6]    &58997.15492630&Apertif&     $348.87\pm0.22$   & [6]    &59407.59936   &CHIME  &   $348.8165\pm 0.0093$& [1]    \\
  58931.51122577&Apertif&     $348.88\pm0.18$   & [6]    &58997.23883623&Apertif&     $348.24\pm0.25$   & [6]    &59440.50726   &CHIME  &   $348.731 \pm 0.043$ & [1]    \\
  58931.54877968&Apertif&     $349.02\pm0.59$   & [6]    &58997.26968437&Apertif&     $348.76\pm0.25$   & [6]    &59440.51585   &CHIME  &   $349.062 \pm 0.012$ & [1]    \\
  58931.56964778&Apertif&     $348.70\pm0.97$   & [6]    &58997.35800780&Apertif&     $348.87\pm0.17$   & [6]    &59486.38963   &CHIME  &   $349.79  \pm 0.21$  & [1]    \\
  58932.550594  &uGMRT  &     $349.06\pm0.32$   & [7]    &58997.38837259&Apertif&     $348.75\pm0.18$   & [6]    &59519.30410   &CHIME  &   $349.9   \pm 0.19$  & [1]    \\
  58932.604069  &uGMRT  &     $348.82\pm0.10$   & [7]    &58998.15708057&Apertif&     $348.69\pm0.26$   & [6]    &59519.30470   &CHIME  &   $348.967 \pm 0.023$ & [1]    \\
  58932.612024  &uGMRT  &     $348.90\pm0.86$   & [7]    &59013.69287   &CHIME  &   $348.955 \pm 0.031$ & [1]    &59519.30909   &CHIME  &   $348.862 \pm 0.014$ & [1]    \\
  58949.53491816&LOFAR  &     $349.03\pm0.11$   & [6]    &59014.68533   &CHIME  &   $348.7917\pm 0.0055$& [1]    &59521.28871   &CHIME  &   $348.94  \pm 0.19$  & [1]    \\
  58949.63987585&LOFAR  &     $348.94\pm0.08$   & [6]    &59031.13727   &uGMRT  &     $349.5\pm0.1$     & [7]    &59569.16266   &CHIME  &   $349.05  \pm 0.16$  & [1]    \\
  58950.52919335&LOFAR  &     $349.02\pm0.08$   & [6]    &59095.01258701&Apertif&     $348.21\pm0.42$   & [6]    &59570.15498   &CHIME  &   $348.901 \pm 0.027$ & [1]    \\
  58950.54130169&LOFAR  &     $349.41\pm0.03$   & [6]    &              &       &                       &        &                      &                       &        \\
  \noalign{\smallskip}\hline
  \end{tabular}
  \begin{flushleft}
                Ref:[1] \cite{Mckinven2023}; [2] \cite{Marcote2020}; [3] \cite{Chawla2020}; [4] \cite{Pearlman2020}; [5] \cite{Pilia2020}; [6] \cite{Pastor-Marazuela2021}; [7] \cite{Marthi2020}; \cite{Pleunis2021}.
  \end{flushleft}
\end{table*}

\subsection{The statistical properties of DM}\label{subsect.2.1}

\begin{figure}
\centering
  \includegraphics[width=\columnwidth]{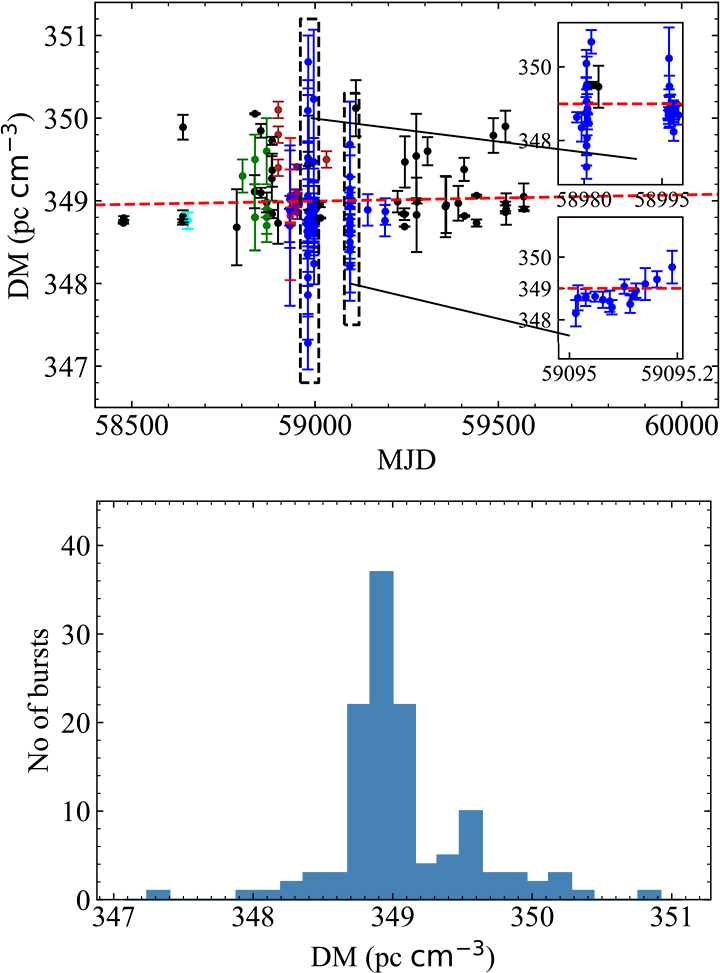}
  \caption{MJD-DM relation (upper) and histogram distribution of DM (bottom) with the 0.16 bin size. In the upper panel, the black, cyan, blue, green, red, brown and purple dots are the observational data from CHIME, EVN, Apertif, GBT, SRT, uGMRT (650 MHz) and LOFAR, respectively. And the red dashed line in the upper panel is the linear fitting result. }\label{DM1}
\end{figure}

 In this subsection, we give out the statistical properties of DMs of FRB 180916 observed by different telescopes. Their arrival times, telescope names, DMs, and references are listed in Table \ref{tb1}.

Since all bursts observed by LOFAR with the given DMs are around 160 MHz \citep{Pastor-Marazuela2021,Pleunis2021}, we take the 160 MHz as a center frequency of bursts observed by LOFAR in the following discussions.
The DM variation of the burst ($\Delta \rm DM \sim 21 \, pc \, cm^{-3}$) at MJD 58883.02020163 is significantly higher than that of other bursts detected by CHIME/FRB \citep{Mckinven2023}.
However, the DM variations of other repeaters are within the range of $\rm 20 \, pc \, cm^{-3}$ \citep{Li2021b,Xu2022}. 
This means that the abnormal bursts reported by \citealt{Pearlman2020} may have a distinctly different origin or de-dispersion algorithm, we will ignore these bursts here.

Figure \ref{DM1} shows the MJD-DM relationship as well as a histogram of DMs. We also give out a linear fitting of the DMs and the slope is $0.028(0.12) \, \rm pc \, cm^{-3} \, yr^{-1}$, indicating that the DM increases over time. The Kolmogorov-Smirnov (K-S) test from scipy.stats.kstest is used to examine the Gaussian property of DMs; their p-value (0.0012) is less than the significant level of 0.05. To perform a K-S test on the random sampling DMs of all data, we take the error bar of each DM observation as the standard deviation and sample 1000 points using a Gaussian random distribution. We discover that the p-value is also less than the 0.05 level of significance. The total DMs are not Gaussian distribution but may have multiple components. 

\begin{figure}
\centering
  \includegraphics[width=\columnwidth]{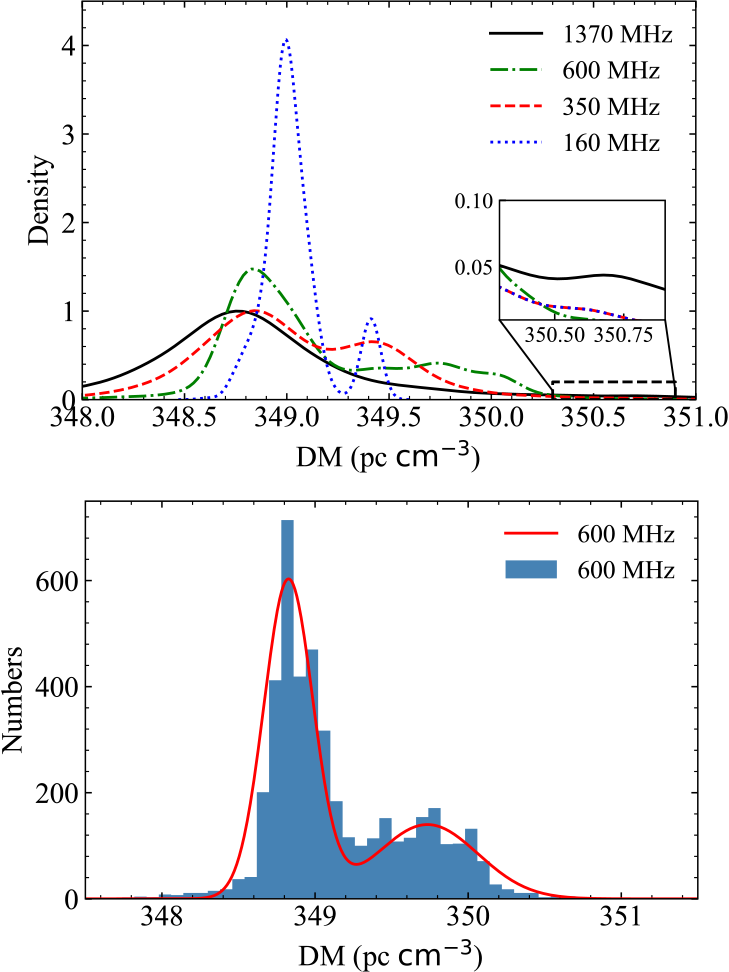}
  \caption{ (Upper panel) KDEs of these DMs at different frequencies. Each DM is sampled 1000 points by using a Gaussian random distribution with the error bar of each DM observation as the standard deviation. (Bottom panel) Histogram of DMs at 600 MHz. The red solid line is fitted by two-component Gaussians, and the Adjusted R-Square is 0.93, the peaks of the Gaussians are at $348.826 \, \rm pc \, cm^{-3}$ and $349.734 \, \rm pc \, cm^{-3}$ respectively.}
  \label{KED}
\end{figure}

To reveal the real distributions, the distribution density of DM with weighted error and different frequencies have been used to estimate the distributions of DMs \citep{Yusifov2004}. Using this method, we find a relatively large difference between the estimated and actual DM, which may be due to a small number of DM samples. As a result, the upper panel of Figure~\ref{KED} depicts the kernel density estimation (KDE) of randomly sampled DMs at different frequencies. The sub-plot demonstrates the KDEs of DMs between $350.3 \, \rm pc\, cm^{-3}$ and $350.9 \, \rm pc\, cm^{-3}$. According to Figure~\ref{KED}, DMs for three frequencies have two main peaks, but the correct distribution for 600 MHz may have multiple peaks. The histogram of DMs for 600 MHz is shown in the bottom panel of Figure ~\ref{KED}. It also shows that the DMs have two main peaks, thus we take the peak DM around $349.734 \, \rm pc \, cm^{-3}$  at this frequency as the right peak. We use the KDE approach for DMs to evaluate two DM peaks, which are repeated 1000 times for each frequency. The mean value of peaking DMs is approximated for each peak. Their standard deviation is set as an error. Table~\ref{peak} finally shows the peaking DMs at four frequencies.
The right peaking DM at 1370 MHz appears more than 250 times but exists, causing a significantly higher inaccuracy in Table \ref{peak}, whereas the other peaking DMs appear 1000 times. Thus, the DM of the left peaks decreases with the frequency increases, but the reverse is true for the right ones.

\begin{table*}
  \centering
  \caption{Two peaking DMs at different frequencies, which are based on one thousand times of KDEs. The subscript `L' denotes the DMs at the left peaks, `R' denotes the right one.}\label{peak}
  \begin{center}
  \begin{tabular}{lcccc}
  \hline\hline\noalign{\smallskip}
  Center frequency                                   & 160 MHz       & 350 MHz   & 600 MHz   & 1370 MHz  \\
  \hline\noalign{\smallskip}
  $\rm DM_{\rm obs, L}$ ($\rm pc \, cm^{-3}$)        & 348.988(1.9)       & 348.854(10.2)       & 348.820(6.9)   & 348.755(12.6)   \\
  $\rm DM_{\rm obs, R}$ ($\rm pc \, cm^{-3}$)        &349.410(2.0)       & 349.453(12.3)       &349.742(8.1)   &  350.436(110.5)   \\
  $\rm \Delta DM_{\rm obs}$ ($\rm pc \, cm^{-3}$)    & 0.422(2.8)         &  0.599(16.0)         & 0.992(10.6)    & 1.774(111.2)     \\
  \noalign{\smallskip}\hline\hline
  \end{tabular}
  \end{center}
\end{table*}

\subsection{The properties of waiting time}\label{subsect.2.2}

The waiting time is calculated by two adjacent bursts within an observational campaign, thus the bursts are taken into account for data from EVN \citep{Marcote2020}, Apertif \citep{Pastor-Marazuela2021}, GBT \citep{Chawla2020}, SRT \citep{Pilia2020}, uGMRT \citep{Marthi2020,Pleunis2021}, LOFAR \citep{Pastor-Marazuela2021,Pleunis2021}, and CHIME. 
Considering the lack of pulse width, fluence, and peak flux density in \citealt{Mckinven2023}, we will use the data given by \citealt{Amiri2020} to maintain consistency in the following discussions.

\begin{figure}
\centering
  \includegraphics[width=\columnwidth]{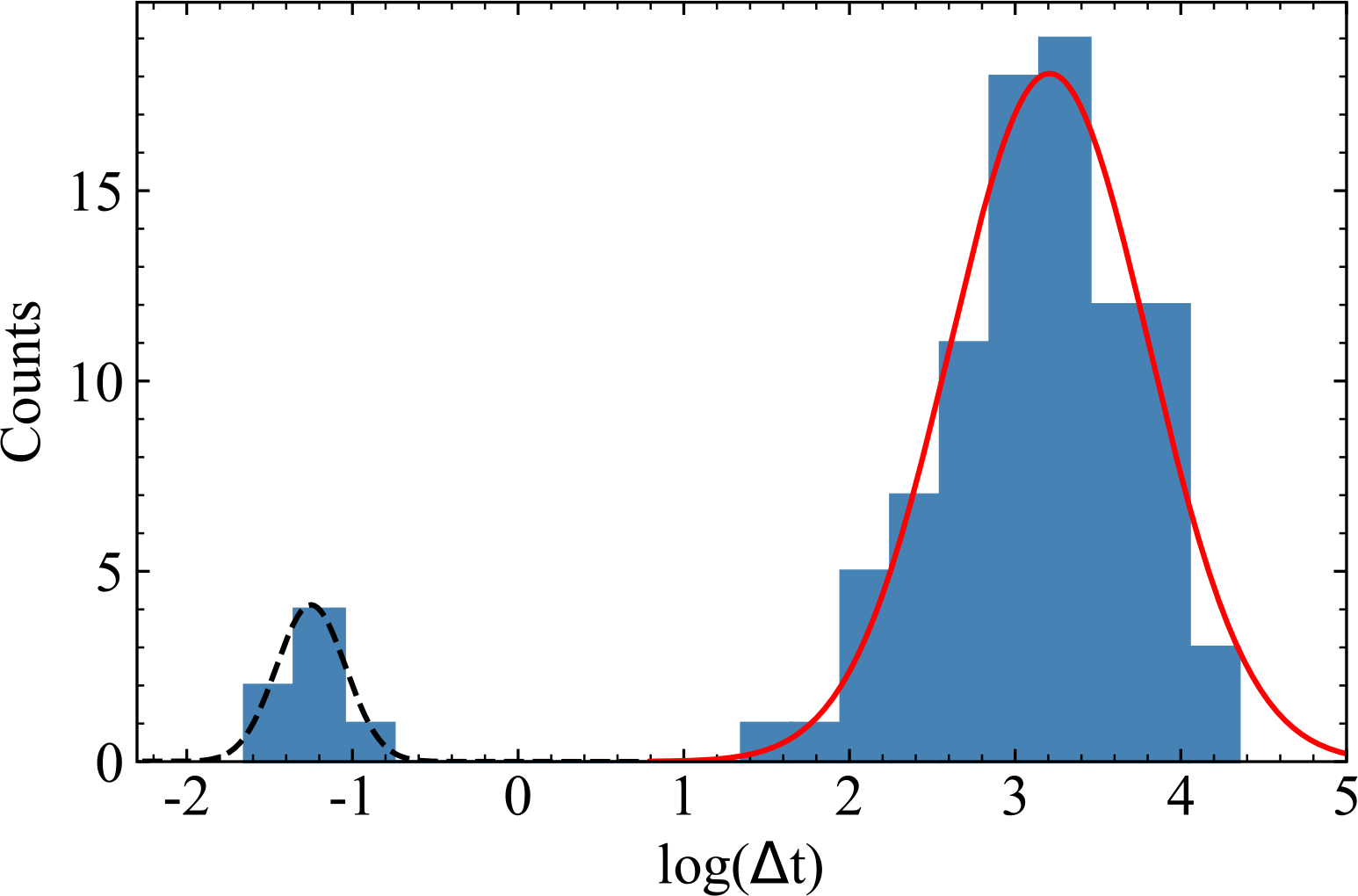}
  \caption{Histogram for the distribution of the waiting time in units of the second. The bin size is 0.3, and the red solid and black dash lines are the fitting curves of the right and left parts of Gaussian functions, respectively.}\label{WT}
\end{figure}

Since the bursts of the repeater appear random at observable frequency bandwidth and an occasional time in the phase window, the histogram of waiting times in the logarithm coordinate is given in Figure \ref{WT}, which shows a typical bimodal distribution.
The waiting times of most bursts are in the range of $31.352 \rm \,s$ to $11397.10608 \rm \,s$.
In the right part, 89 waiting times are well fitted by the Gaussian function with a reduced Chi-Square of $\sim2.737$ and an Adjusted R-Square of $\sim 0.918$, their peaking value is at $\Delta t = 1612.91266 \rm\, s$.
We used the K-S test to examine the right part of the waiting time and found that the p-value (0.974) is greater than the 0.05 significance level.
Thus the waiting times on the right part cannot reject a Gaussian distribution.
We are also aware that CHIME/FRB only has an exposure of 12 min or $\sim$40 min each time \citep{Amiri2020,Chawla2020}, and other telescopes usually have few-hour exposures each time, so it is difficult to see pairs longer than the CHIME/FRB telescope observation length, which may lead to the decrease in counts greater than $10^3$ s (17 min). However,
our result can be consistent with the burst rate ($\sim$1.8 bursts hr$^{-1}$) of the repeater \citep{Amiri2020,Chawla2020}. In the left part, seven waiting times vary around tens of milliseconds, appearing in the 300$\sim$800 MHz and frequency-dependent.
The mean waiting times at different frequencies are $37.4 \, \rm ms$ ($350 \, \rm MHz$), $52.7 \, \rm ms$ ($600 \, \rm MHz$), and $86.4\, \rm ms$ ($650 \, \rm MHz$), respectively.
We also use the Gaussian function to fit the left part and obtain the peaking value at $\Delta t = 0.05622 \, \rm s$.
The Anderson-Darling (A-D) test gives out their statistic value (0.421) and critical value (0.742). These results suggest that waiting time follows a Log-normal distribution. We thought the decrease in the waiting time less than 10 ms may be related to the width of the bursts and the definition of individual bursts, which may be caused by the different physical properties discussed in the following.

\begin{figure*}
\centering
  \includegraphics[width=160mm]{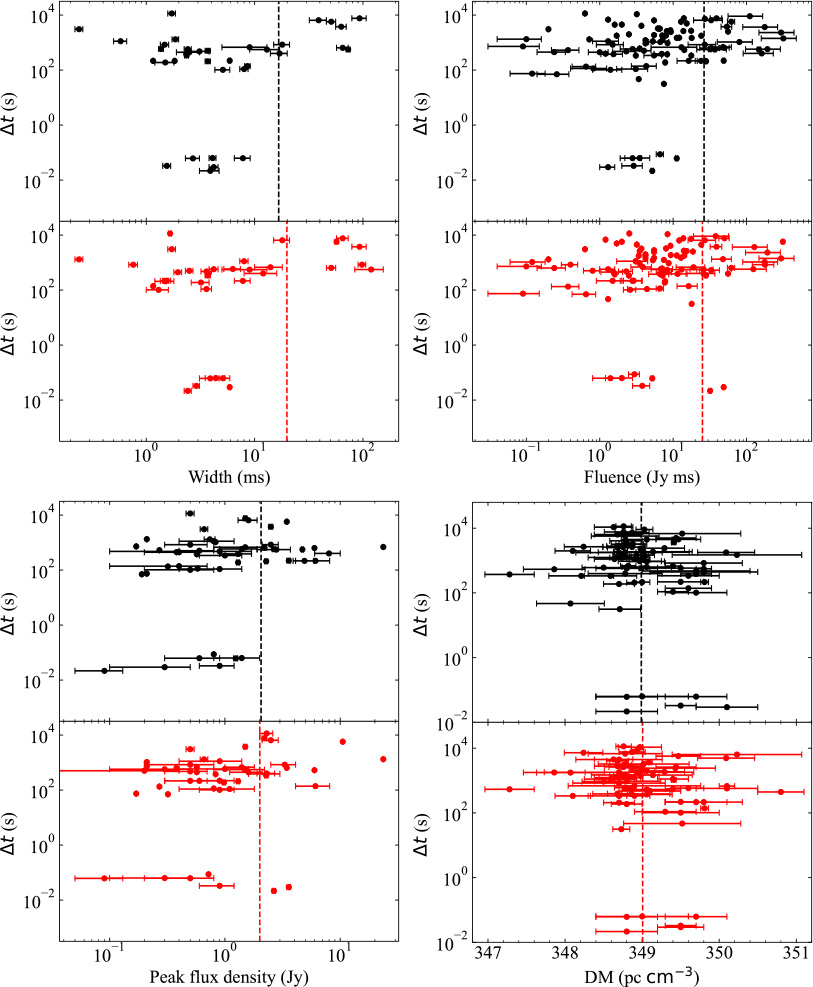}
  \caption{Waiting time vs. pulse width (the left upper panel), fluence (the right upper panel), peak flux density (the left bottom panels) and DM (the right bottom panel) of FRB bursts. The black dots represent the data of the previous bursts, the red dots are for these from the subsequent bursts, and the dashed lines are the mean values.}\label{WT_X}
\end{figure*}

\begin{figure}
\centering
  \includegraphics[width=\columnwidth]{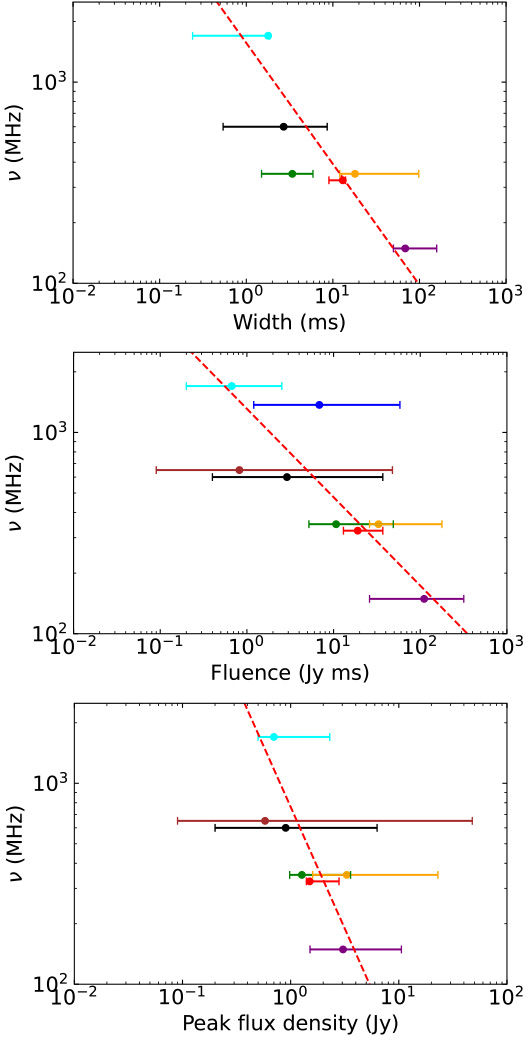}
  \caption{Observational center frequency vs. median for pulse width (the upper panel), fluency (the middle panel) and peak flux density (the bottom panel). The black, cyan, blue, green, red, brown, orange and purple dots are the data from CHIME, EVN, Apertif, GBT, SRT, uGMRT ($650 \rm \, MHz$), uGMRT ($350 \, \rm MHz$) and LOFAR, respectively. The error bars are the minimum and maximum values of parameters, the red dashed lines are the linear fitting in the log-log coordinate.}\label{Frequency_x}
\end{figure}

The accumulated energy of an FRB burst for the next one depends on the duration of two consecutive bursts.
The possible correlation between the waiting time and the burst
intensity could appear in the preceding or subsequent bursts, which
is presented in Figure \ref{WT_X}. The DMs of subsequent bursts have a higher mean value than that in previous bursts. The mean fluence and peak flux density for the subsequent bursts are provided with relatively lower values, and the mean pulse width of previous bursts is narrow, they suggest that the width may have opposite properties with the mean fluence and peak flux density. It is easily found that all plots are scattered data points, so these parameters
have no obvious correlation with the waiting time.
We also take the observational data in Figure \ref{WT_X} to examine the correlations between these parameters and the frequency. In Figure \ref{Frequency_x}, the pulse width, fluence, and peak flux density show anti-correlative with the frequency in the Log-Log coordinate. The
variations of mean values for the three parameters of the previous and
subsequent bursts are irrelevant to the frequency. Coupled with the results of 
previous statistical analysis on the DMs and waiting time, some external mechanisms and the effects in the propagation path
are a more probable contribution to the repeater.

\section{The lensing effects on FRB 180916.J0158+65}\label{sect.3}

In this section, we will discuss multiple images caused by the lens. Furthermore, the delay time difference of multiple images within the lens model will be calculated through the statistical results of DMs.

For the axially symmetric gravitational lensing in the vacuum, the light rays propagate along the null geodesic curve and are converged by the gravitational field.
Since the light rays from the two sides of the lens could have a propagation path difference, an observer may detect two frequency-independent images with different arrival times and magnifications, e.g., the review by \citealt{Bisnovatyi-Kogan2017}.
The delay time difference between two images should be smaller than $15 \rm \, ms$ for primordial black holes \citep{Munoz2016}.
The two images with different propagation paths could also undergo different non-magnetized cold plasma. Thus the dispersive delay difference depends on the contribution of their DM difference ($\Delta \rm DM(\nu)$) and is simplified as
\begin{eqnarray}\label{dispersive}
\Delta t_{\rm dis}  = 4.15 {\rm \, ms} \,\frac{\Delta \rm DM(\nu)}{\nu^2_{\rm GHz}},
\end{eqnarray}
where $\nu_{\rm GHz} = \nu/\rm GHz$ is the frequency of the image in the unit of $\rm GHz$.
From Table \ref{peak}, $\Delta t_{\rm dis}$ at $600 \, \rm MHz$, $350 \, \rm MHz$ and $160 \, \rm MHz$ are $ 11.44 \, \rm ms$, $20.29 \, \rm ms$ and $68.41 \, \rm ms$, respectively.
Coupled with the gravitational lensing effects, they are still incompatible with the results of Figure \ref{WT}.
Some other mechanisms may exist in the propagation path of the repeater.

The radio signals passing through the plasma lens can diverge as multiple images with different propagation paths. The DMs of images are the frequency-dependence and can present the multiple-components DM distributions in the observation (our previous paper, i.e., \citealt{Wang2022b}). 
Moreover, the delay time difference between two images depends on the variations of their DMs and propagation paths.

Thus, the higher-order effects of DM variations in the theoretical prediction should be introduced to the delay time differences, which are roughly expressed as
\begin{eqnarray}\label{tot}
\Delta t_{\rm PL}  = \chi + 4.15 {\rm \, ms} \,\frac{\Delta \rm DM}{\nu^2_{\rm GHz}} - b \frac{\Delta \rm DM^2}{\nu^4_{\rm GHz}},
\end{eqnarray}
where $\chi$ is the geometric delay time difference contributed by the repeater itself \citep{Wang2019} and some plasma clouds with relation $ \overset{n}{\underset{i = 1}{\sum}} {\rm \Delta DM}_i \propto \nu^2$ \citep{Tuntsov2021}, the second term in the right side of equation is the classical dispersion relation, the third term is the delay time difference due to the geometric effects, and $b$ is a free parameter.
From the results at 350 MHz and 600 MHz in Section 2, we can get $ \chi = 52.51 \, \rm ms$ and $b =1.48\, \rm ms$ through Equation (\ref{tot}).
Consequently, the delay time differences at the 160 MHz and 1370 MHz are $-281.38 \rm \, ms$ and $55.11 \rm \, ms$, respectively.

\begin{table*}
  \centering
  \caption{Two peaking DMs, DM differences ($\Delta\rm DM_{\rm G}$) and the delay time
differences ($\Delta t_{\rm G}$) at different center frequencies calculated from Equation (\ref{tot}). The subscript `G' is for the results from the two-component Gaussian function.}\label{Gauss_func}
  \begin{center}
  \begin{tabular}{lccccc}
  \hline\hline\noalign{\smallskip}
  Center frequency                             & 160 MHz     & 350 MHz   & 600 MHz   & 650 MHz   & 1370 MHz  \\
  \hline\noalign{\smallskip}
  $\rm DM_{\rm G, L}$ ($\rm pc \, cm^{-3}$)    &  348.919     & 348.847   & 348.805  & 348.799   & 348.742   \\
  $\rm DM_{\rm G, R}$ ($\rm pc \, cm^{-3}$)    & 349.252      & 349.505   & 349.724  &  349.762   & 350.406   \\
  $\rm \Delta DM_{\rm G}$ ($\rm pc \, cm^{-3}$)&  0.333       & 0.658     & 0.919    &  0.963     & 1.664     \\
  $\Delta t_{\rm G}$ ($\rm ms$)                & -144.01      & 32.08    & 53.45    & 54.27     & 55.02     \\
  \noalign{\smallskip}\hline\hline
  \end{tabular}
  \end{center}
\end{table*}

Due to the physical properties of the plasma lens, the frequencies of corresponding DMs should be taken as positive values.
These bursts produced by a single source cannot occupy two peaking DMs while $\nu \rightarrow \infty$. We found the two-component Gaussian function is the best fit for the peaking DMs in the Table \ref{peak} (an Adjusted R-square of $\sim$  0.967).
For the DM distribution, the two peaking DMs at five frequencies, the DM differences ($\Delta \rm DM_{\rm G}$) for each frequency and the delay time differences are given in Table \ref{Gauss_func}.
The delay times differences are decreasing with the decreasing frequency except for the value at 160 MHz.
We found that $\rm DM_{\rm G,R}$ at 160 MHz is relatively lower than $\rm DM_{\rm obs, R}$.
The DMs listed in Table \ref{tb1}, except for the burst at MJD 58950.54130169, are around $348.98 \rm \, pc \, cm^{-3}$ at 160 MHz, and their scattering timescales are around $46\, \rm ms$ at 150 MHz \citep{Pastor-Marazuela2021}.
All these results suggested that the scattering effect may contribute $\sim 0.249 \, \rm pc\, cm^{-3}$ to the DM of burst at MJD 58950.54130169. Furthermore, it can cause a more significant delay time difference than a single burst of $\rm Width \gtrsim 40 \, \rm ms$.
Therefore, the delay time differences and the frequency-dependent DMs are consistent with the plasma lensing model \citep{Wang2022b}.
FRB 180916.J0158+65 may be affected by the plasma lens with the two-component Gaussian model describing DM distributions.

\section{Summary and Discussions}\label{Sect.4}

We investigate the repeating features of FRB 180916.J0158+65
in terms of statistics, especially waiting times and DMs.
The DMs have a precise bimodal distribution at different frequencies. The peaking values of DMs increase with frequency drop for the left distribution but decrease with frequency decrease for the right distribution, demonstrating that repeating bursts at different frequencies could come from different propagation paths \citep{Wang2022b}. This viewpoint is supported by the degree of linear polarization with frequency-dependent characteristics in the repeated FRBs \citep{Feng2022}.
We fit these DM peaks
and find that the two-component Gaussian function
is more appropriate for the DM peaks at different frequencies. 
The waiting time is also a discontinuous bimodal distribution peaking
at 56.22 ms and 1612.91266 s, the mean waiting time in the left
distribution decreases as the frequency decrease. The parameters of the preceding or subsequent bursts, including pulse width, fluence, peak flux density and DM, do not correlate with the waiting times. Considering the preceding and subsequent burst parameters, external effects may contribute to the repeater.

Based on the statistics for observed bursts of FRB 180916.J0158+65, the repeater suffered from lensing effects in the propagation path has been examined.
The delay time difference between two images produced by the gravitational lensing is inconsistent with the waiting time.
The higher-order term from the plasma lensing effects can be consistent with the left peak distribution of the waiting times and the variations of DMs, except at 160 MHz.
As a relatively clean line-of-sight path at 160 MHz \citep{Pleunis2021}, the scattering effect may highly contribute to the burst at the right distribution of DMs. 
The frequency-dependent DM may contribute to the properties of near-source plasma and the intervening galaxy, such as the supernova remnants around the source, the pulsar wind nebulae produced by the source, HII regions, and galactic halo \citep{Yang2017,Prochaska2019,Er2022}.
The multi-frequency observations of some repeaters are an important way to reveal their propagating effects and the plasma lensing model. 

FRB 180916.J0158+65 may originate from the binary system.
For the NS$-$white dwarf (WD) scenario \citep{Gu2016,Gu2020}, the waiting times vary from $ 100 \,$ to $\rm 1.59392 \times 10^4 \, \rm s$ when the Eddington limit are considered ($\dot{M} \lesssim 10^{18} \, \rm g ~s^{-1}$).
But the energy of a burst from the gravitational potential energy of the accreted material is incompatible with the observation \citep{Frank2002}.
In the NS$-$asteroid belt collision scenario \citep{Geng2015,Dai2016,Dai2020}, the typical energy of a burst ($10^{36} - 10^{38} \, \rm erg$) depends on the gravitational energy of the asteroid \citep{Dai2016,Smallwood2019}, but the long time to the next collision ($\sim 0.8 \, \rm h$) \citep{Dai2016} is also inconsistent with the observations of the repeater.

For the ``cosmic comb'' model \citep{Zhang2017}, 
the stellar wind in the NS$-$NS binary scenario or NS$-$black hole binary scenario cannot be strong enough to explain the properties of the repeater FRB 180916.J0158+65 \citep{Du2021}.
According to the maximal energy ($\sim 6\times 10^{38} \, \rm erg$) of the repeater, the dipole magnetic field of NS at $1 \,\rm GHz$ requires $B_{\rm d} \gtrsim 4.18 \times 10^{13} \rm \, G$  \citep{Kumar2017,Metzger2019}. For an high mass X-ray binary (HMXB) system, the model requires the pulsar with the spin period of $P \gtrsim 1.09 \, \rm s \, B_{\rm d,12}^{1/3}$\citep{Walter2015}.
That means the highly magnetized NS in HMXB may be of origin of FRB 180916.J0158+65.
In addition, the multiple bursts in this model can be produced in the pulsar's magnetosphere \citep{Wang2019,Levkov2022}.
The interval time between two bursts is $\lesssim 18.70 \, {\rm ms}$ when taken $P=23.5~\rm{s}$ and the radial radius within the light cylinder radius, which suggests that the zero-order term of FRB 180916.J0158+65 given by Equation (\ref{tot}) may be mainly contributed by other effects in the propagation path \citep{Tuntsov2021}.
Till now, 3 in 24 repeaters, including FRB 180916.J0158+65, FRB 121102 and FRB 20201124A, have some similar statistical properties \citep{Li2021b,Xu2022},
FRB 20201124A may reside in a magnetar/Be star binary \citep{Wang2022a}.
\textbf
This type of repeater may be related to a highly magnetized NS in the HMXB.

\section*{Acknowledgements}

This work is supported in part by the Opening Foundation of Xinjiang Key Laboratory (No. 2021D04016), the National Natural Science Foundation of China (Nos. 12033001, 12288102, 11833003, 12273028, 12041304), the Natural Science Foundation of Xinjiang Uygur Autonomous Region (No. 2022D01A363), the Major Science and Technology Program of Xinjiang Uygur Autonomous Region (No. 2022A03013-1, 2022A03013-3), the Chinese Academy of Sciences (CAS) “Light of West China” Program (No. 2018-XBQNXZ-B-025), and the special research assistance project of the CAS. X. Z. would like to thank Qing-min Li for her fruitful suggestions.

\section*{Data availability}

Observational data used in this paper are quoted from the cited works.
Additional data generated from computations can be made available
upon reasonable request.



\bibliographystyle{mnras}
\bibliography{ref.bib} 

\begin{thebibliography}{}
\makeatletter
\relax
\def\mn@urlcharsother{\let\do\@makeother \do\$\do\&\do\#\do\^\do\_\do\%\do\~}
\def\mn@doi{\begingroup\mn@urlcharsother \@ifnextchar [ {\mn@doi@}
  {\mn@doi@[]}}
\def\mn@doi@[#1]#2{\def\@tempa{#1}\ifx\@tempa\@empty \href
  {http://dx.doi.org/#2} {doi:#2}\else \href {http://dx.doi.org/#2} {#1}\fi
  \endgroup}
\def\mn@eprint#1#2{\mn@eprint@#1:#2::\@nil}
\def\mn@eprint@arXiv#1{\href {http://arxiv.org/abs/#1} {{\tt arXiv:#1}}}
\def\mn@eprint@dblp#1{\href {http://dblp.uni-trier.de/rec/bibtex/#1.xml}
  {dblp:#1}}
\def\mn@eprint@#1:#2:#3:#4\@nil{\def\@tempa {#1}\def\@tempb {#2}\def\@tempc
  {#3}\ifx \@tempc \@empty \let \@tempc \@tempb \let \@tempb \@tempa \fi \ifx
  \@tempb \@empty \def\@tempb {arXiv}\fi \@ifundefined
  {mn@eprint@\@tempb}{\@tempb:\@tempc}{\expandafter \expandafter \csname
  mn@eprint@\@tempb\endcsname \expandafter{\@tempc}}}

\bibitem[\protect\citeauthoryear{{Abbate} et~al.,}{{Abbate}
  et~al.}{2020}]{Abbate2020}
{Abbate} F.,  et~al., 2020, \mn@doi [\mnras] {10.1093/mnras/staa2510}, \href
  {https://ui.adsabs.harvard.edu/abs/2020MNRAS.498..875A} {498, 875}

\bibitem[\protect\citeauthoryear{{Amiri} et~al.,}{{Amiri}
  et~al.}{2020}]{Amiri2020}
{Amiri} M.,  et~al., 2020, \mn@doi [\nat] {10.1038/s41586-020-2398-2}, \href
  {https://ui.adsabs.harvard.edu/abs/2020Natur.582..351C} {582, 351}

\bibitem[\protect\citeauthoryear{{Amiri} et~al.,}{{Amiri}
  et~al.}{2021}]{Amiri2021}
{Amiri} M.,  et~al., 2021, \mn@doi [\apjs] {10.3847/1538-4365/ac33ab}, \href
  {https://ui.adsabs.harvard.edu/abs/2021ApJS..257...59C} {257, 59}

\bibitem[\protect\citeauthoryear{{Andersen} et~al.,}{{Andersen}
  et~al.}{2019}]{Andersen2019}
{Andersen} B.~C.,  et~al., 2019, \mn@doi [\apjl] {10.3847/2041-8213/ab4a80},
  \href {https://ui.adsabs.harvard.edu/abs/2019ApJ...885L..24C} {885, L24}

\bibitem[\protect\citeauthoryear{{Andersen} et~al.,}{{Andersen}
  et~al.}{2020}]{Andersen2020}
{Andersen} B.~C.,  et~al., 2020, \mn@doi [\nat] {10.1038/s41586-020-2863-y},
  \href {https://ui.adsabs.harvard.edu/abs/2020Natur.587...54C} {587, 54}

\bibitem[\protect\citeauthoryear{{Bethapudi}, {Spitler}, {Main}, {Li}  \&
  {Wharton}}{{Bethapudi} et~al.}{2022}]{Bethapud2022}
{Bethapudi} S.,  {Spitler} L.~G.,  {Main} R.~A.,  {Li} D.~Z.,   {Wharton}
  R.~S.,  2022, arXiv e-prints, \href
  {https://ui.adsabs.harvard.edu/abs/2022arXiv220713669B} {p. arXiv:2207.13669}

\bibitem[\protect\citeauthoryear{{Bisnovatyi-Kogan} \&
  {Tsupko}}{{Bisnovatyi-Kogan} \& {Tsupko}}{2017}]{Bisnovatyi-Kogan2017}
{Bisnovatyi-Kogan} G.,  {Tsupko} O.,  2017, \mn@doi [Universe]
  {10.3390/universe3030057}, \href
  {https://ui.adsabs.harvard.edu/abs/2017Univ....3...57B} {3, 57}

\bibitem[\protect\citeauthoryear{{Chamma}, {Rajabi}, {Wyenberg}, {Mathews}  \&
  {Houde}}{{Chamma} et~al.}{2021}]{Chamma2021}
{Chamma} M.~A.,  {Rajabi} F.,  {Wyenberg} C.~M.,  {Mathews} A.,   {Houde} M.,
  2021, \mn@doi [\mnras] {10.1093/mnras/stab2070}, \href
  {https://ui.adsabs.harvard.edu/abs/2021MNRAS.507..246C} {507, 246}

\bibitem[\protect\citeauthoryear{{Chawla} et~al.,}{{Chawla}
  et~al.}{2020}]{Chawla2020}
{Chawla} P.,  et~al., 2020, \mn@doi [\apjl] {10.3847/2041-8213/ab96bf}, \href
  {https://ui.adsabs.harvard.edu/abs/2020ApJ...896L..41C} {896, L41}

\bibitem[\protect\citeauthoryear{{Cheng}, {Zhang}  \& {Wang}}{{Cheng}
  et~al.}{2020}]{Cheng2020}
{Cheng} Y.,  {Zhang} G.~Q.,   {Wang} F.~Y.,  2020, \mn@doi [\mnras]
  {10.1093/mnras/stz3085}, \href
  {https://ui.adsabs.harvard.edu/abs/2020MNRAS.491.1498C} {491, 1498}

\bibitem[\protect\citeauthoryear{{Cordes} \& {Lazio}}{{Cordes} \&
  {Lazio}}{2002}]{Cordes2002}
{Cordes} J.~M.,  {Lazio} T.~J.~W.,  2002, arXiv e-prints, \href
  {https://ui.adsabs.harvard.edu/abs/2002astro.ph..7156C} {pp
  astro--ph/0207156}

\bibitem[\protect\citeauthoryear{{Dai} \& {Zhong}}{{Dai} \&
  {Zhong}}{2020}]{Dai2020}
{Dai} Z.~G.,  {Zhong} S.~Q.,  2020, \mn@doi [\apjl] {10.3847/2041-8213/ab8f2d},
  \href {https://ui.adsabs.harvard.edu/abs/2020ApJ...895L...1D} {895, L1}

\bibitem[\protect\citeauthoryear{{Dai}, {Wang}, {Wu}  \& {Huang}}{{Dai}
  et~al.}{2016}]{Dai2016}
{Dai} Z.~G.,  {Wang} J.~S.,  {Wu} X.~F.,   {Huang} Y.~F.,  2016, \mn@doi [\apj]
  {10.3847/0004-637X/829/1/27}, \href
  {https://ui.adsabs.harvard.edu/abs/2016ApJ...829...27D} {829, 27}

\bibitem[\protect\citeauthoryear{{Du}, {Wang}, {Wu}  \& {Xu}}{{Du}
  et~al.}{2021}]{Du2021}
{Du} S.,  {Wang} W.,  {Wu} X.,   {Xu} R.,  2021, \mn@doi [\mnras]
  {10.1093/mnras/staa3527}, \href
  {https://ui.adsabs.harvard.edu/abs/2021MNRAS.500.4678D} {500, 4678}

\bibitem[\protect\citeauthoryear{{Er} \& {Mao}}{{Er} \& {Mao}}{2022}]{Er2022}
{Er} X.,  {Mao} S.,  2022, \mn@doi [\mnras] {10.1093/mnras/stac2323}, \href
  {https://ui.adsabs.harvard.edu/abs/2022MNRAS.516.2218E} {516, 2218}

\bibitem[\protect\citeauthoryear{{Feng} et~al.,}{{Feng}
  et~al.}{2022}]{Feng2022}
{Feng} Y.,  et~al., 2022, \mn@doi [Science] {10.1126/science.abl7759}, \href
  {https://ui.adsabs.harvard.edu/abs/2022Sci...375.1266F} {375, 1266}

\bibitem[\protect\citeauthoryear{{Frank}, {King}  \& {Raine}}{{Frank}
  et~al.}{2002}]{Frank2002}
{Frank} J.,  {King} A.,   {Raine} D.~J.,  2002, {Accretion Power in
  Astrophysics: Third Edition}

\bibitem[\protect\citeauthoryear{{Geng} \& {Huang}}{{Geng} \&
  {Huang}}{2015}]{Geng2015}
{Geng} J.~J.,  {Huang} Y.~F.,  2015, \mn@doi [\apj]
  {10.1088/0004-637X/809/1/24}, \href
  {https://ui.adsabs.harvard.edu/abs/2015ApJ...809...24G} {809, 24}

\bibitem[\protect\citeauthoryear{{Geng}, {Li}  \& {Huang}}{{Geng}
  et~al.}{2021}]{Geng2021}
{Geng} J.,  {Li} B.,   {Huang} Y.,  2021, \mn@doi [The Innovation]
  {10.1016/j.xinn.2021.100152}, \href
  {https://ui.adsabs.harvard.edu/abs/2021Innov...200152G} {2, 100152}

\bibitem[\protect\citeauthoryear{{Geyer} et~al.,}{{Geyer}
  et~al.}{2021}]{Geyer2021}
{Geyer} M.,  et~al., 2021, \mn@doi [\mnras] {10.1093/mnras/stab1501}, \href
  {https://ui.adsabs.harvard.edu/abs/2021MNRAS.505.4468G} {505, 4468}

\bibitem[\protect\citeauthoryear{{Gu}, {Dong}, {Liu}, {Ma}  \& {Wang}}{{Gu}
  et~al.}{2016}]{Gu2016}
{Gu} W.-M.,  {Dong} Y.-Z.,  {Liu} T.,  {Ma} R.,   {Wang} J.,  2016, \mn@doi
  [\apjl] {10.3847/2041-8205/823/2/L28}, \href
  {https://ui.adsabs.harvard.edu/abs/2016ApJ...823L..28G} {823, L28}

\bibitem[\protect\citeauthoryear{{Gu}, {Yi}  \& {Liu}}{{Gu}
  et~al.}{2020}]{Gu2020}
{Gu} W.-M.,  {Yi} T.,   {Liu} T.,  2020, \mn@doi [\mnras]
  {10.1093/mnras/staa1914}, \href
  {https://ui.adsabs.harvard.edu/abs/2020MNRAS.497.1543G} {497, 1543}

\bibitem[\protect\citeauthoryear{{Kuiack}, {Wijers}, {Rowlinson}, {Shulevski},
  {Huizinga}, {Molenaar}  \& {Prasad}}{{Kuiack} et~al.}{2020}]{Kuiack2020}
{Kuiack} M.,  {Wijers} R. A.~M.~J.,  {Rowlinson} A.,  {Shulevski} A.,
  {Huizinga} F.,  {Molenaar} G.,   {Prasad} P.,  2020, \mn@doi [\mnras]
  {10.1093/mnras/staa1996}, \href
  {https://ui.adsabs.harvard.edu/abs/2020MNRAS.497..846K} {497, 846}

\bibitem[\protect\citeauthoryear{{Kumar}, {Lu}  \& {Bhattacharya}}{{Kumar}
  et~al.}{2017}]{Kumar2017}
{Kumar} P.,  {Lu} W.,   {Bhattacharya} M.,  2017, \mn@doi [\mnras]
  {10.1093/mnras/stx665}, \href
  {https://ui.adsabs.harvard.edu/abs/2017MNRAS.468.2726K} {468, 2726}

\bibitem[\protect\citeauthoryear{{Kumar}, {Shannon}, {Lower}, {Bhandari},
  {Deller}, {Flynn}  \& {Keane}}{{Kumar} et~al.}{2022}]{Kumar2022}
{Kumar} P.,  {Shannon} R.~M.,  {Lower} M.~E.,  {Bhandari} S.,  {Deller} A.~T.,
  {Flynn} C.,   {Keane} E.~F.,  2022, \mn@doi [\mnras] {10.1093/mnras/stac683},
  \href {https://ui.adsabs.harvard.edu/abs/2022MNRAS.512.3400K} {512, 3400}

\bibitem[\protect\citeauthoryear{{Kurban} et~al.,}{{Kurban}
  et~al.}{2022}]{Kurban2022}
{Kurban} A.,  et~al., 2022, \mn@doi [\apj] {10.3847/1538-4357/ac558f}, \href
  {https://ui.adsabs.harvard.edu/abs/2022ApJ...928...94K} {928, 94}

\bibitem[\protect\citeauthoryear{{Levkov}, {Panin}  \& {Tkachev}}{{Levkov}
  et~al.}{2022}]{Levkov2022}
{Levkov} D.~G.,  {Panin} A.~G.,   {Tkachev} I.~I.,  2022, \mn@doi [\apj]
  {10.3847/1538-4357/ac3250}, \href
  {https://ui.adsabs.harvard.edu/abs/2022ApJ...925..109L} {925, 109}

\bibitem[\protect\citeauthoryear{{Li}, {Li}, {Zhang}, {Geng}, {Song}, {Huang}
  \& {Yang}}{{Li} et~al.}{2019}]{Li2019}
{Li} B.,  {Li} L.-B.,  {Zhang} Z.-B.,  {Geng} J.-J.,  {Song} L.-M.,  {Huang}
  Y.-F.,   {Yang} Y.-P.,  2019, arXiv e-prints, \href
  {https://ui.adsabs.harvard.edu/abs/2019arXiv190103484L} {p. arXiv:1901.03484}

\bibitem[\protect\citeauthoryear{{Li} et~al.,}{{Li} et~al.}{2021a}]{Li2021a}
{Li} C.~K.,  et~al., 2021a, \mn@doi [Nature Astronomy]
  {10.1038/s41550-021-01302-6}, \href
  {https://ui.adsabs.harvard.edu/abs/2021NatAs...5..378L} {5, 378}

\bibitem[\protect\citeauthoryear{{Li} et~al.,}{{Li} et~al.}{2021b}]{Li2021b}
{Li} D.,  et~al., 2021b, \mn@doi [\nat] {10.1038/s41586-021-03878-5}, \href
  {https://ui.adsabs.harvard.edu/abs/2021Natur.598..267L} {598, 267}

\bibitem[\protect\citeauthoryear{{Li}, {Li}, {Zhong}, {Xia}, {Xie}, {Wang}  \&
  {Dai}}{{Li} et~al.}{2022}]{Li2022}
{Li} L.,  {Li} Q.-C.,  {Zhong} S.-Q.,  {Xia} J.,  {Xie} L.,  {Wang} F.-Y.,
  {Dai} Z.-G.,  2022, \mn@doi [\apj] {10.3847/1538-4357/ac5d5a}, \href
  {https://ui.adsabs.harvard.edu/abs/2022ApJ...929..139L} {929, 139}

\bibitem[\protect\citeauthoryear{{Luo} et~al.,}{{Luo} et~al.}{2020}]{Luo2020}
{Luo} R.,  et~al., 2020, \mn@doi [\nat] {10.1038/s41586-020-2827-2}, \href
  {https://ui.adsabs.harvard.edu/abs/2020Natur.586..693L} {586, 693}

\bibitem[\protect\citeauthoryear{{Mao} et~al.,}{{Mao} et~al.}{2022}]{Mao2022}
{Mao} J.-W.,  et~al., 2022, \mn@doi [Research in Astronomy and Astrophysics]
  {10.1088/1674-4527/ac6797}, \href
  {https://ui.adsabs.harvard.edu/abs/2022RAA....22f5006M} {22, 065006}

\bibitem[\protect\citeauthoryear{{Marcote} et~al.,}{{Marcote}
  et~al.}{2020}]{Marcote2020}
{Marcote} B.,  et~al., 2020, \mn@doi [\nat] {10.1038/s41586-019-1866-z}, \href
  {https://ui.adsabs.harvard.edu/abs/2020Natur.577..190M} {577, 190}

\bibitem[\protect\citeauthoryear{{Marthi}, {Gautam}, {Li}, {Lin}, {Main},
  {Naidu}, {Pen}  \& {Wharton}}{{Marthi} et~al.}{2020}]{Marthi2020}
{Marthi} V.~R.,  {Gautam} T.,  {Li} D.~Z.,  {Lin} H.~H.,  {Main} R.~A.,
  {Naidu} A.,  {Pen} U.~L.,   {Wharton} R.~S.,  2020, \mn@doi [\mnras]
  {10.1093/mnrasl/slaa148}, \href
  {https://ui.adsabs.harvard.edu/abs/2020MNRAS.499L..16M} {499, L16}

\bibitem[\protect\citeauthoryear{{Mckinven} et~al.,}{{Mckinven}
  et~al.}{2023}]{Mckinven2023}
{Mckinven} R.,  et~al., 2023, \mn@doi [\apj] {10.3847/1538-4357/acc65f}, \href
  {https://ui.adsabs.harvard.edu/abs/2023ApJ...950...12M} {950, 12}

\bibitem[\protect\citeauthoryear{{Metzger}, {Margalit}  \& {Sironi}}{{Metzger}
  et~al.}{2019}]{Metzger2019}
{Metzger} B.~D.,  {Margalit} B.,   {Sironi} L.,  2019, \mn@doi [\mnras]
  {10.1093/mnras/stz700}, \href
  {https://ui.adsabs.harvard.edu/abs/2019MNRAS.485.4091M} {485, 4091}

\bibitem[\protect\citeauthoryear{{Mu{\~n}oz}, {Kovetz}, {Dai}  \&
  {Kamionkowski}}{{Mu{\~n}oz} et~al.}{2016}]{Munoz2016}
{Mu{\~n}oz} J.~B.,  {Kovetz} E.~D.,  {Dai} L.,   {Kamionkowski} M.,  2016,
  \mn@doi [\prl] {10.1103/PhysRevLett.117.091301}, \href
  {https://ui.adsabs.harvard.edu/abs/2016PhRvL.117i1301M} {117, 091301}

\bibitem[\protect\citeauthoryear{{Pastor-Marazuela} et~al.,}{{Pastor-Marazuela}
  et~al.}{2021}]{Pastor-Marazuela2021}
{Pastor-Marazuela} I.,  et~al., 2021, \mn@doi [\nat]
  {10.1038/s41586-021-03724-8}, \href
  {https://ui.adsabs.harvard.edu/abs/2021Natur.596..505P} {596, 505}

\bibitem[\protect\citeauthoryear{{Pearlman}, {Majid}, {Prince}, {Nimmo},
  {Hessels}, {Naudet}  \& {Kocz}}{{Pearlman} et~al.}{2020}]{Pearlman2020}
{Pearlman} A.~B.,  {Majid} W.~A.,  {Prince} T.~A.,  {Nimmo} K.,  {Hessels} J.
  W.~T.,  {Naudet} C.~J.,   {Kocz} J.,  2020, \mn@doi [\apjl]
  {10.3847/2041-8213/abca31}, \href
  {https://ui.adsabs.harvard.edu/abs/2020ApJ...905L..27P} {905, L27}

\bibitem[\protect\citeauthoryear{{Petroff} et~al.,}{{Petroff}
  et~al.}{2016}]{Petroff2016}
{Petroff} E.,  et~al., 2016, \mn@doi [\pasa] {10.1017/pasa.2016.35}, \href
  {https://ui.adsabs.harvard.edu/abs/2016PASA...33...45P} {33, e045}

\bibitem[\protect\citeauthoryear{{Pilia} et~al.,}{{Pilia}
  et~al.}{2020}]{Pilia2020}
{Pilia} M.,  et~al., 2020, \mn@doi [\apjl] {10.3847/2041-8213/ab96c0}, \href
  {https://ui.adsabs.harvard.edu/abs/2020ApJ...896L..40P} {896, L40}

\bibitem[\protect\citeauthoryear{{Platts}, {Weltman}, {Walters}, {Tendulkar},
  {Gordin}  \& {Kandhai}}{{Platts} et~al.}{2019}]{Platts2019}
{Platts} E.,  {Weltman} A.,  {Walters} A.,  {Tendulkar} S.~P.,  {Gordin}
  J.~E.~B.,   {Kandhai} S.,  2019, \mn@doi [\physrep]
  {10.1016/j.physrep.2019.06.003}, \href
  {https://ui.adsabs.harvard.edu/abs/2019PhR...821....1P} {821, 1}

\bibitem[\protect\citeauthoryear{{Platts} et~al.,}{{Platts}
  et~al.}{2021}]{Platts2021}
{Platts} E.,  et~al., 2021, \mn@doi [\mnras] {10.1093/mnras/stab1544}, \href
  {https://ui.adsabs.harvard.edu/abs/2021MNRAS.505.3041P} {505, 3041}

\bibitem[\protect\citeauthoryear{{Pleunis} et~al.,}{{Pleunis}
  et~al.}{2021}]{Pleunis2021}
{Pleunis} Z.,  et~al., 2021, \mn@doi [\apjl] {10.3847/2041-8213/abec72}, \href
  {https://ui.adsabs.harvard.edu/abs/2021ApJ...911L...3P} {911, L3}

\bibitem[\protect\citeauthoryear{{Prochaska} et~al.,}{{Prochaska}
  et~al.}{2019}]{Prochaska2019}
{Prochaska} J.~X.,  et~al., 2019, \mn@doi [Science] {10.1126/science.aay0073},
  \href {https://ui.adsabs.harvard.edu/abs/2019Sci...366..231P} {366, 231}

\bibitem[\protect\citeauthoryear{{Rajwade} et~al.,}{{Rajwade}
  et~al.}{2020}]{Rajwade2020}
{Rajwade} K.~M.,  et~al., 2020, \mn@doi [\mnras] {10.1093/mnras/staa1237},
  \href {https://ui.adsabs.harvard.edu/abs/2020MNRAS.495.3551R} {495, 3551}

\bibitem[\protect\citeauthoryear{{Ravi} et~al.,}{{Ravi}
  et~al.}{2022}]{Ravi2022}
{Ravi} V.,  et~al., 2022, \mn@doi [\mnras] {10.1093/mnras/stac465}, \href
  {https://ui.adsabs.harvard.edu/abs/2022MNRAS.513..982R} {513, 982}

\bibitem[\protect\citeauthoryear{{Smallwood}, {Martin}  \& {Zhang}}{{Smallwood}
  et~al.}{2019}]{Smallwood2019}
{Smallwood} J.~L.,  {Martin} R.~G.,   {Zhang} B.,  2019, \mn@doi [\mnras]
  {10.1093/mnras/stz483}, \href
  {https://ui.adsabs.harvard.edu/abs/2019MNRAS.485.1367S} {485, 1367}

\bibitem[\protect\citeauthoryear{{Tendulkar} et~al.,}{{Tendulkar}
  et~al.}{2017}]{Tendulkar2017}
{Tendulkar} S.~P.,  et~al., 2017, \mn@doi [\apjl] {10.3847/2041-8213/834/2/L7},
  \href {https://ui.adsabs.harvard.edu/abs/2017ApJ...834L...7T} {834, L7}

\bibitem[\protect\citeauthoryear{{Tendulkar} et~al.,}{{Tendulkar}
  et~al.}{2021}]{Tendulkar2021}
{Tendulkar} S.~P.,  et~al., 2021, \mn@doi [\apjl] {10.3847/2041-8213/abdb38},
  \href {https://ui.adsabs.harvard.edu/abs/2021ApJ...908L..12T} {908, L12}

\bibitem[\protect\citeauthoryear{{Thornton} et~al.,}{{Thornton}
  et~al.}{2013}]{Thornton2013}
{Thornton} D.,  et~al., 2013, \mn@doi [Science] {10.1126/science.1236789},
  \href {https://ui.adsabs.harvard.edu/abs/2013Sci...341...53T} {341, 53}

\bibitem[\protect\citeauthoryear{{Tuntsov}, {Pen}  \& {Walker}}{{Tuntsov}
  et~al.}{2021}]{Tuntsov2021}
{Tuntsov} A.,  {Pen} U.-L.,   {Walker} M.,  2021, arXiv e-prints, \href
  {https://ui.adsabs.harvard.edu/abs/2021arXiv210713549T} {p. arXiv:2107.13549}

\bibitem[\protect\citeauthoryear{{Walter}, {Lutovinov}, {Bozzo}  \&
  {Tsygankov}}{{Walter} et~al.}{2015}]{Walter2015}
{Walter} R.,  {Lutovinov} A.~A.,  {Bozzo} E.,   {Tsygankov} S.~S.,  2015,
  \mn@doi [\aapr] {10.1007/s00159-015-0082-6}, \href
  {https://ui.adsabs.harvard.edu/abs/2015A&ARv..23....2W} {23, 2}

\bibitem[\protect\citeauthoryear{{Wang}, {Zhang}, {Chen}  \& {Xu}}{{Wang}
  et~al.}{2019}]{Wang2019}
{Wang} W.,  {Zhang} B.,  {Chen} X.,   {Xu} R.,  2019, \mn@doi [\apjl]
  {10.3847/2041-8213/ab1aab}, \href
  {https://ui.adsabs.harvard.edu/abs/2019ApJ...876L..15W} {876, L15}

\bibitem[\protect\citeauthoryear{{Wang}, {Zhang}, {Dai}  \& {Cheng}}{{Wang}
  et~al.}{2022a}]{Wang2022a}
{Wang} F.~Y.,  {Zhang} G.~Q.,  {Dai} Z.~G.,   {Cheng} K.~S.,  2022a, \mn@doi
  [Nature Communications] {10.1038/s41467-022-31923-y}, \href
  {https://ui.adsabs.harvard.edu/abs/2022NatCo..13.4382W} {13, 4382}

\bibitem[\protect\citeauthoryear{{Wang}, {Wen}, {Yuen}, {Wang}, {Yuan}  \&
  {Zhou}}{{Wang} et~al.}{2022b}]{Wang2022b}
{Wang} Y.-B.,  {Wen} Z.-G.,  {Yuen} R.,  {Wang} N.,  {Yuan} J.-P.,   {Zhou} X.,
   2022b, \mn@doi [Research in Astronomy and Astrophysics]
  {10.1088/1674-4527/ac6aad}, \href
  {https://ui.adsabs.harvard.edu/abs/2022RAA....22f5017W} {22, 065017}

\bibitem[\protect\citeauthoryear{{Xu} et~al.,}{{Xu} et~al.}{2022}]{Xu2022}
{Xu} H.,  et~al., 2022, \mn@doi [\nat] {10.1038/s41586-022-05071-8}, \href
  {https://ui.adsabs.harvard.edu/abs/2022Natur.609..685X} {609, 685}

\bibitem[\protect\citeauthoryear{{Yang} \& {Zhang}}{{Yang} \&
  {Zhang}}{2017}]{Yang2017}
{Yang} Y.-P.,  {Zhang} B.,  2017, \mn@doi [\apj] {10.3847/1538-4357/aa8721},
  \href {https://ui.adsabs.harvard.edu/abs/2017ApJ...847...22Y} {847, 22}

\bibitem[\protect\citeauthoryear{{Yao}, {Manchester}  \& {Wang}}{{Yao}
  et~al.}{2017}]{Yao2017}
{Yao} J.~M.,  {Manchester} R.~N.,   {Wang} N.,  2017, \mn@doi [\apj]
  {10.3847/1538-4357/835/1/29}, \href
  {https://ui.adsabs.harvard.edu/abs/2017ApJ...835...29Y} {835, 29}

\bibitem[\protect\citeauthoryear{{Younes} et~al.,}{{Younes}
  et~al.}{2020}]{Younes2020}
{Younes} G.,  et~al., 2020, \mn@doi [\apjl] {10.3847/2041-8213/abc94c}, \href
  {https://ui.adsabs.harvard.edu/abs/2020ApJ...904L..21Y} {904, L21}

\bibitem[\protect\citeauthoryear{{Yusifov} \& {K{\"u}{\c{c}}{\"u}k}}{{Yusifov}
  \& {K{\"u}{\c{c}}{\"u}k}}{2004}]{Yusifov2004}
{Yusifov} I.,  {K{\"u}{\c{c}}{\"u}k} I.,  2004, \mn@doi [\aap]
  {10.1051/0004-6361:20040152}, \href
  {https://ui.adsabs.harvard.edu/abs/2004A&A...422..545Y} {422, 545}

\bibitem[\protect\citeauthoryear{{Zhang}}{{Zhang}}{2017}]{Zhang2017}
{Zhang} B.,  2017, \mn@doi [\apjl] {10.3847/2041-8213/aa5ded}, \href
  {https://ui.adsabs.harvard.edu/abs/2017ApJ...836L..32Z} {836, L32}

\makeatother
\end{thebibliography}


\bsp	
\label{lastpage}
\end{document}